\documentstyle[12pt]{article}

\catcode`\@=11 \@addtoreset{equation}{section}\catcode`\@=12
\begin{document}
\thispagestyle{empty}

\begin{center}
               RUSSIAN GRAVITATIONAL ASSOCIATION\\
               CENTER FOR SURFACE AND VACUUM RESEARCH\\
               DEPARTMENT OF FUNDAMENTAL INTERACTIONS AND METROLOGY\\
\end{center}
\vskip 4ex
\begin{flushright}                 RGA-CSVR-002/94\\
                                   gr-qc/9403064
\end{flushright}
\vskip 45mm

\begin{center}
{\bf
MULTIDIMENSIONAL COSMOLOGY WITH $m$-COMPONENT PERFECT FLUID}
\vskip 5mm
{\bf V. D. Ivashchuk and V. N. Melnikov}\\
\vskip 5mm
     {\em Centre for Surface and Vacuum Research,\\
     8 Kravchenko str., Moscow, 117331, Russia}\\
     e-mail: mel@cvsi.uucp.free.msk.su\\
\vskip 60mm

             Moscow 1994
\end{center}
\pagebreak

\setcounter{page}{1}
\begin{center}
{\bf
MULTIDIMENSIONAL COSMOLOGY WITH $m$-COMPONENT PERFECT FLUID}
\footnote{This is an extended version of the talk presented by the authors
at the 8th Russian Gravitational Conference (Pushchino, 25-28 May
1993)}.                                                                 \\
 \vskip 1ex
{\bf V. D. Ivashchuk and V. N. Melnikov}\\
 \vskip 1ex

\end{center}
ABSTRACT
 \vskip 1ex

A cosmological model describing the evolution of $n$ Einstein
spaces $(n>1)$ with  $m$-component perfect-fluid matter is considered. When
all spaces are Ricci-flat and for any $\alpha$-th component the pressures
in all spaces are proportional to the density:
$p_{i}^{(\alpha)} = (1- h_{i}^{(\alpha)}) \rho^{(\alpha)}$,
$h_{i}^{(\alpha)}$ = const, the Einstein and Wheeler-DeWitt equations are
integrated in the cases: i) $m=1$, for all $h_{i}^{(\alpha)}$;
ii) $m > 1$, for some special sets of $h_{i}^{(\alpha)}$. For $m=1$
the quantum wormhole solutions are also obtained. \\
\vskip 2ex

\vskip 2ex

\section{Introduction. The model}

Multidimensional cosmology (see, for example [1-21] and references
therein) is a very interesting object of investigations both from
physical and mathematical points of view.

Last decade the interest in multidimensional cosmology
was stimulated mainly by Kaluza-Klein and superstring paradigmas [22,23].
The "realistic" multidimensional cosmological models appeared mainly
in a context of some unifications theories. Certainly, it is quite natural
to believe that the Entire Universe is multidimensional and we live in
a some sort of a (3+1)- dimensional layer, that is Our Universe.
Of course, at first stage we should try to understand the structure of
our 3-dimensional crude (dense) matter and the formation of Our Universe.
But it seems to be very likely that at some stage of our development
it will be just impossible to describe our (3+1)-dimensional layer
(Our Universe) out of touch with other (multidimensional) layers and
domains.

A large variety of multidimensional cosmological
models is described by pseudo-Euclidean Toda-like systems [19] (see formula
(1.10) below). These systems are not well studied yet. We note, that
the Euclidean Toda-like systems are more or less well studied [24-28]
(at least for certain sets of parameters, associated with
finite-dimensional Lie algebras or affine Lie algebras). There is also a
criterion of integrability by quadrature (algebraic integrability)
for these (Euclidean) systems established by Adler and van Moerbeke [28].
Nevertheless, there are some indications that cosmological models may
contain rather rich mathematical structures. For example, a self-dual
reduction of the Bianchi-IX cosmology [29] lead us to the Halphen
system of ordinary differential equations [30]. This system may be
integrated in terms of modular forms [31] and is connected with a
certain integrable reduction of the self-dual Yang-Mills equation [32]
(with the infinite-dimensional group $SDiffSU(2)$). Another example
is connected with the Kaluza-Klein dyon solution from [33]. The field
equations for a spherically-symmetric Kaluza-Klein dyon in 5-dimensions
were reduced in [33] to an open (Euclidean) Toda lattice with three
points. Certainly, this problem may be formulated in terms of an
appropriate cosmological model described by a pseudo-Euclidean Toda-
like Lagrangian. So, we lead to an interesting nontrivial example of
an integrable cosmological model.

In this paper we consider a cosmological model describing
the evolution of $n$ Einstein spaces in the presence of $m$-component
perfect-fluid matter. The metric of the model
\begin{equation} g=-exp[2{\gamma}(t)]dt \otimes dt +
\sum_{i=1}^{n} exp[2{x^{i}}(t)]g^{(i)},
\end{equation}
is defined on the  manifold \begin{equation}
M = R \times M_{1} \times \ldots \times M_{n}, \end{equation}
where the manifold $M_{i}$ with the metric $g^{(i)}$ is an
Einstein space of dimension $N_{i}$, i.e. \begin{equation}
{R_{m_{i}n_{i}}}[g^{(i)}] = \lambda^{i} g^{(i)}_{m_{i}n_{i}},
\end{equation}
$i = 1, \ldots ,n $; $n \geq 2$.
The energy-momentum tensor is adopted in the following form
\begin{eqnarray}
&& T^{M}_{N} = \sum_{\alpha =1}^{m} T^{M (\alpha)}_{N}, \\
&&(T^{M (\alpha)}_{N})= diag(-{\rho^{(\alpha)}}(t),
 {p_{1}^{(\alpha)}}(t) \delta^{m_{1}}_{k_{1}},
\ldots , {p^{(\alpha)}_{n}}(t) \delta^{m_{n}}_{k_{n}}).
\end{eqnarray}
$\alpha =1, \ldots ,m$, with the conservation law constraints imposed:
\begin{equation}
\bigtriangledown_{M} T^{M (\alpha)}_{N}=0 \end{equation}
$\alpha=1, \ldots ,m-1$.
The Einstein equations
\begin{equation}
R^{M}_{N}-\frac{1}{2}\delta^{M}_{N}R=\kappa^{2}T^{M}_{N}
\end{equation}
($\kappa^{2}$ is gravitational constant) imply
$\bigtriangledown_{M} T^{M}_{N}=0$ and consequently
$\bigtriangledown_{M} T^{M (m)}_{N}=0$.

We suppose that for any $\alpha$-th component of matter the
pressures in all spaces are proportional to the density
\begin{equation}
{p_{i}^{(\alpha)}}(t) = (1-{h_{i}^{(\alpha)}}(x(t))){\rho^{(\alpha)}}(t),
\end{equation}
where \begin{equation}
{h_{i}^{(\alpha)}}(x) = \frac{1}{N_{i}}
\frac{\partial}{\partial x^{i}}{\Phi^{(\alpha)}}(x),
\end{equation}
$i=1, \ldots ,n$, where ${\Phi^{(\alpha)}}(x)$ is a smooth function
on $R^{n}$, $\alpha=1, \ldots , m$.

In Sec. 2 the Einstein equations (1.7) for the model are reduced to
the equations of motion for some Lagrange system with the energy
constraint $E=0$ imposed. When $m=1$ and all spaces are Ricci-flat
($\lambda^{i} =0$ in (1.3), $i=1, \ldots ,n$) such reduction was
performed previously in [9].

In Sec. 3 we consider the Einstein equations, when all spaces are
Ricci-flat and $h_{i}^{(\alpha)} = const$, $i=1, \ldots ,n$,
$\alpha=1, \ldots ,m$.
In this case we deal with pseudo-Euclidean Toda-like system with the
Lagrangian
\begin{equation} L_{A} = \frac{1}{2} G_{ij}\dot{x}^{i}\dot{x}^{j}-
 \sum_{\alpha=1}^{m} \kappa^{2}A^{(\alpha)}\exp(u_{i}^{(\alpha)} x^{i}),
\end{equation}
where $sign(G_{ij})=(-,+, \ldots, +)$ [14,15],
$u_{i}^{(\alpha)} = N_{i}h_{i}^{(\alpha)}$ and $A^{(\alpha)} = const$
$i=1, \ldots ,n$, $\alpha=1, \ldots, m$. The Einstein equations are
integrated in the following cases: 1) $m=1$;
2) $n=2$, $m \geq 2$, $A^{(\alpha)} \neq 0$,
$u^{(\alpha)} - u^{(1)} = b^{(\alpha)} u$, $\alpha=1, \ldots, m$, where
$u^{2} =G^{ij}u_{i}u_{j}=0$, $u \neq 0$; 3) $u^{(\alpha)} = b^{(\alpha)} u$,
$u^{2} < 0$, $A^{(\alpha)} > 0$, $\alpha=1, \ldots, m$.

In Sec. 4 the Wheeler-DeWitt (WDW) equation for the model is considered.
When all spaces are Ricci-flat the WDW equation is integrated in all listed
above cases. For $m=1$ the solution of the WDW equation, satisfying
so-called quantum wormhole boundary conditions [34], are obtained.

\section{The equations of motion}

The non-zero components of the Ricci-tensor for the metric
(1.1) are following
\begin{equation}
R_{00}=- \sum_{i=1}^{n} N_{i}[ \ddot{x}^{i} - \dot{\gamma} \dot{x}^{i}
+ (\dot{x}^{i})^{2}], \end{equation} \begin{equation}
R_{m_{i}n_{i}}=g_{m_{i}n_{i}}^{(i)} [\lambda^{i} +\exp(2x^{i}-2\gamma)
(\ddot{x}^{i}+\dot{x}^{i}(\sum_{i=1}^{n}N_{i}\dot{x}^{i}-\dot{\gamma}))],
\end{equation}
$i=1,\ldots,n$.

We put \begin{equation}
\gamma = \gamma_{0} \equiv \sum_{i=1}^{n} N_{i}x^{i} \end{equation}
in (1.1) (the harmonic time is used). Then it follows from  (2.1) and (2.2)
that the Einstein equations (1.7) for the metric (1.1) with $\gamma$
from (2.3) and the energy-momentum tensor from (1.4), (1.5) are
equivalent to the following set of equations
\begin{equation}
\frac{1}{2}G_{ij}\dot{x}^{i}\dot{x}^{j}+ V_{c} + \kappa^{2}
\sum_{\alpha=1}^{m} \rho^{(\alpha)} \exp(2\gamma_{0})= 0,
\end{equation}
\begin{equation}
\lambda^{i}+ \ddot{x}^{i}\exp(2x^{i}-2\gamma_{0})=
\kappa^{2}\exp(2x^{i}) \sum_{\alpha=1}^{m} [p_{i}^{(\alpha)} +
(D-2)^{-1}(\rho^{(\alpha)} -\sum_{j=1}^{n}N_{j}p_{j}^{(\alpha)})],
\end{equation} $i=1, \ldots ,n$. Here
\begin{equation}
G_{ij}=N_{i}\delta_{ij}- N_{i}N_{j} \end{equation}
are the components of the minisuperspace metric,
\begin{equation}
V_{c} =-\frac{1}{2}\sum_{i=1}^{n}\lambda^{i}N_{i}exp(-2x^{i}+2\gamma_{0})
\end{equation} is the potential and $D \equiv dimM =1+ \sum_{i=1}^{n}
N_{i}$.

The conservation law constraint (1.6) for $\alpha \in \{1,...,m\}$ reads
\begin{equation}
\dot{\rho}^{(\alpha)}+
\sum_{i=1}^{n}N_{i}\dot{x}^{i}(\rho^{(\alpha)} + p_{i}^{(\alpha)})=0.
\end{equation}
We impose the conditions of state in the form (1.8), (1.9). Then
eq. (2.8) gives  \begin{equation}
{\rho^{(\alpha)}}(t)=A^{(\alpha)}
exp[-2N_{i}{x^{i}}(t) +{\Phi^{(\alpha)}}({x}(t))],
\end{equation}
where $A^{(\alpha)}=const$ and eqs. (2.4), (2.5) may be written in the
following manner \begin{equation}
\frac{1}{2}G_{ij}\dot{x}^{i}\dot{x}^{j}+
V + \kappa^{2} \sum_{\alpha=1}^{m} A^{(\alpha)} \exp\Phi^{(\alpha)}= 0,
\end{equation}
\begin{equation}
\lambda^{i}+ \ddot{x}^{i}\exp(2x^{i}-2\gamma_{0})=
- \kappa^{2} \sum_{\alpha=1}^{m} u^{i}_{(\alpha)} A^{(\alpha)}
\exp(2x^{i}-2\gamma_{0}+\Phi^{(\alpha)}),
\end{equation} $i=1, \ldots ,n$. In  (2.11) we denote
\begin{equation}
u_{i}^{(\alpha)} \equiv N_{i}h_{i}^{(\alpha)}= \partial_{i}\Phi^{(\alpha)},
\qquad u^{i}_{(\alpha)}= G^{ij}u_{j}^{(\alpha)}, \end{equation}
where [15]
\begin{equation}
G^{ij} = \frac{\delta^{ij}}{N_{i}}+\frac{1}{2-D} \end{equation}
are the components of the matrix inverse to the matrix $(G_{ij})$
(2.6).

It is not difficult to verify that equations (2.11) are equivalent
to the Lagrange equations for the Lagrangian
\begin{equation}
L = \frac{1}{2} G_{ij}\dot{x}^{i}\dot{x}^{j}- V
\end{equation}
where \begin{equation}
V ={V}(x)= {V_{c}}(x) +
\sum_{\alpha=1}^{m} \kappa^{2}A^{(\alpha)} \exp[{\Phi^{(\alpha)}}(x)].
\end{equation} Eq. (2.10) is the zero-energy constraint
\begin{equation}
E = \frac{1}{2} G_{ij}\dot{x}^{i}\dot{x}^{j}+ V = 0.
\end{equation}

Remark 1. In terms of 1-forms $u^{(\alpha)} = u_{i}^{(\alpha)} dx^{i}$,
the relations (1.9) read: $u^{(\alpha)} = d \Phi^{(\alpha)}$,
$\alpha=1, \ldots , m$. In this case
\begin{equation}
du^{(\alpha)} =0,
\end{equation}
$\alpha=1, \ldots , m$. The set of eqs. (2.17) (on $R^{n}$) is equivalent
to (1.9). An open problem is to generalize the considered here formalism
for the following cases: a) $du^{(\alpha)} \neq 0$ for some
$\alpha \in  \{ 1, \ldots , m \}$; b) $du^{(\alpha)} = 0$ for all
$\alpha=1, \ldots , m$, but $u^{(\alpha)}$ are defined on an
open submanifold $\Omega \in R^{n}$ with the non-trivial
cohomology group ${H^{1}}(\Omega,R) \neq 0$.

Using eqs. (2.1) and (2.2), it is not difficult to verify that the
Einstein equations (1.7) for the metric (1.1) and the energy-momentum
tensor from (1.4), (1.5), (1.8), (1.9) are equivalent to the Lagrange
equations for the following degenerate Lagrangian (see also [15])
\begin{equation}
L = \frac{1}{2} \exp(- \gamma + {\gamma_{0}}(x)) G_{ij}
\dot{x}^{i}\dot{x}^{j}- \exp( \gamma - {\gamma_{0}}(x)) {V}(x)
\end{equation}
($L = {L}(\gamma,x, \dot{x}$)). Fixing the gauge
\begin{equation}
\gamma  = {\gamma_{0}}(x)  -2{f}(x),
\end{equation}
where $f= {f}(x)$ is a smooth function on $R^{n}$,
we get the Lagrangian
\begin{equation}
L_{f} = \frac{1}{2} \exp(2{f}(x)) G_{ij}
\dot{x}^{i}\dot{x}^{j}- \exp(-2{f}(x)) {V}(x).
\end{equation}
For $f=0$ we have the harmonic-time gauge (2.3).
The set of Lagrange equations for the Lagrangian
(2.18) (or equivalently the set of the  Einstein equations)
with $\gamma$ from (2.19) is equivalent to the set of Lagrange
equations for the Lagrangian (2.20) with the energy constraint
imposed
\begin{equation}
E_{f} = \frac{1}{2} \exp(2{f}(x)) G_{ij}
\dot{x}^{i}\dot{x}^{j}+ \exp(-2{f}(x)) {V}(x) = 0.
\end{equation}

Remark 2. We remind that the action of the relativistic particle
of mass $m$, moving in the pseudo-Euclidean background space with
the metric  ${\hat{G}_{ij}}(x)$ has the following form
\begin{equation}
S = \int d\tau [{\hat{G}_{ij}}({x}(\tau))\frac{\dot{x}^{i}\dot{x}^{j}}
{2{e}(\tau)} - \frac{m^{2}}{2}{e}(\tau)],
\end{equation}
where $e={e}(\tau) $ is 1-bein. Comparing (2.18) and (2.22), we
find that for ${V}(x) > 0$ the cosmological model (2.18) is
equivalent to the model of relativistic particle with the mass
$m=1$, moving in the conformally-flat (pseudo-Euclidean) space with
the metric ${\hat{G}_{ij}}(x) =2{V}(x)G_{ij}$ (see also [37]).
In this case $e = 2{V}(x)\exp( \gamma - {\gamma_{0}}(x))$ .
For ${V}(x) < 0$ we have a tachyon. The problem may be  also
reformulated in terms geodesic-flow problem for conformally-flat
metric (this follows from (2.22) or from a more general scheme [43]).

\section{Classical solutions}

Now, we consider the following case: $\lambda^{i}=0$
(all spaces are Ricci-flat), $u_{i}^{(\alpha)} =
N_{i}h_{i}^{(\alpha)}= const$, $i=1, \ldots ,n$.
Then $V_{c} = 0$ and we put $\Phi^{(\alpha)} =
u_{i}^{(\alpha)} x^{i}$ in (2.15).
In this case the Lagrangian (2.14) has the form (1.10).

Remark 3. The curvature induced term $V_{c}$ (2.7) may be generated
in the framework of the model with the Ricci-flat spaces $M_{i}$
by the addition of $n$ new components of the perfect fluid with
$u_{i}^{(k)} = 2N_{i}- 2 \delta_{i}^{k}$ and $\kappa^{2}A^{(k)}=
-\lambda^{k}N_{k}/2$, $i,k = 1, \ldots ,n$. The introduction of the
cosmological constant $\Lambda$ into the model is equaivalent to
the addition of a new component with
$u_{i}^{(n+1)} = 2N_{i}$ and $\kappa^{2}A^{(n+1)}= \Lambda$.

{\bf 3.1 One-component matter}

We consider the case $m=1$, $A^{(1)} =A \neq 0$. We denote
$h^{(1)}_{i} =h_{i}$, $u^{(1)}_{i} = u_{i} =N_{i} h_{i}$.

We remind [14, 15] that the minisuperspace metric \begin{equation}
G = G_{ij}dx^{i}\otimes dx^{i}  \end{equation}
has pseudo-Euclidean signature $(-,+, \ldots ,+)$, i.e. there exist
a linear transformation \begin{equation}
z^{a}=V^{a}_{i}x^{i}, \end{equation}
diagonalizing the minisuperpace metric (3.1) \begin{equation}
G= \eta_{ab}dz^{a} \otimes dz^{b}=
 -dz^{0} \otimes dz^{0} + \sum_{i=1}^{n-1}dz^{i}\otimes dz^{i},
\end{equation} where \begin{equation}
(\eta_{ab})=(\eta^{ab}) \equiv diag(-1,+1, \ldots ,+1),
\end{equation} $a,b = 0, \ldots ,n-1$.

Proposition 1. For any $u=(u_{i}) \in R^{n}$, $u \neq 0$,
there exists a (nondegenegate) $n \times n$ matrix
$(V^{a}_{i})$ such that \begin{equation}
\eta_{ab}V^{a}_{i}V^{b}_{j}=G_{ij} \end{equation}
and
a) $V^{0}_{i}=u_{i}/\sqrt{-u^{2}}$, for  $u^{2}<0$;
b) $V^{1}_{i}=u_{i}/\sqrt{u^{2}}$, for  $u^{2}>0$;
c) $V^{0}_{i}+V^{1}_{i}= u_{i}$, for  $u^{2}=0$;

Here and below ($u = (u_{i}) = (N_{i}h_{i})$)
\begin{equation} u^{2} \equiv
u_{i}u^{i}= G^{ij}u_{i}u_{j}= \sum_{i=1}^{n}N_{i}(h_{i})^{2}
+\frac{1}{2-D}(\sum_{i=1}^{n}N_{i}h_{i})^{2}. \end{equation}
(We note that in notations of [14] $u^{2}={\acute{\Delta}}(h)/
(2-D)$.)

This proposition follows from the fact that $<u,v> \equiv G^{ij}u_{i}v_{j}$
is bilinear symmetric 2-form of signature $(-,+, \ldots ,+)$ and the
following quite obvious

Proposition 2. Let $v \in E=R^{n}$, $n \geq 2$, and $<.,.> :
E \times E \longrightarrow R$ is a bilinear symmetric 2-form of
signature $(-,+, \ldots,+)$. Then there exists a basis
${v^{0}, \ldots ,v^{n-1}}$ in $E$, such that $<v^{a},v^{b}>=
\eta^{ab}$ and a) $v=v^{0}$, b) $v=v^{1}$, c) $v=v^{0}+v^{1}$,
 in the cases: a) $v^{2} \equiv <v,v>=-1$,
b) $v^{2}=1$, c) $v^{2}=0$ respectively.

Let $u \neq 0$. In $z=(z^{a})$-coordinates (3.2) with the matrix
$(V^{a}_{i})$ from the Proposition 1 the Lagrangian (2.14)
has the following form
\begin{equation}
L_{A} = \frac{1}{2} \eta_{ab} \dot{z}^{a} \dot{z}^{b} - V_{A} =
        -\frac{1}{2}(\dot{z}^{0})^{2} + \sum_{i=1}^{n-1}
           \frac{1}{2}(\dot{z}^{i})^{2} -V_{A},
\end{equation}
where
\begin{eqnarray}
V_{A}&& =\kappa^{2}A \exp(2qz^{0}), \qquad u^{2}<0, \\
&& =\kappa^{2}A \exp(2qz^{1}), \qquad u^{2}>0, \\
&& =\kappa^{2}A \exp(z^{0}+z^{1}), \qquad u^{2}=0,
\end{eqnarray}
is the potential (2.15). Here we denote \begin{equation}
2q \equiv \sqrt{|u^{2}|}.  \end{equation}

The Lagrange equations for the Lagrangian (3.7)
\begin{equation}
\ddot{z}^{a}  = -\eta^{ab}\partial_{b} V_{A}
\end{equation}
with the energy constraint (2.16)
\begin{equation}
E_{A} = \frac{1}{2}\eta_{ab}\dot{z}^{a}\dot{z}^{b}+ V_{A}=0,
\end{equation}
can be easily solved. We present the solutions.

a) For  $u^{2}<0$
\begin{eqnarray}
&&z^{i} = p^{i}t + q^{i}, \qquad          i=1, \ldots , n-1,  \\
&&2qz^{0} = {y}(t),
\end{eqnarray}
where $p^{i}$ and $q^{i}$ are constants and
\begin{eqnarray}
{y}(t)= && \ln[C/D \sinh^{2}(\frac{1}{2}\sqrt{C}(t-t_{0}))], C \neq 0,
D>0, \\ = && \ln[4/D(t-t_{0})^{2}], \qquad C=0, D>0, \\
= && \ln[-C/D \cosh^{2}(\frac{1}{2}\sqrt{C}(t-t_{0}))], C>0, D<0,
\end{eqnarray} Here $t_{0}$ is an arbitrary constant,
$D=-2 u^{2} \kappa^{2} A$, $C = -u^{2}(\vec{p})^{2}$ and
$(\vec{p})^{2} =  \sum_{i=1}^{n-1} (p^{i})^{2}$.

b) For $u^{2} >0$ we have
\begin{eqnarray}
&&z^{i} = p^{i}t + q^{i}, \qquad          i=0,2, \ldots ,n-1 , \\
&&2qz^{1} = {y}(t),
\end{eqnarray}
with $(\vec{p})^{2} = (p^{0})^{2} - \sum_{i=2}^{n-1} (p^{i})^{2}$
in (3.15)-(3.18).

c) $u^{2}=0$, $u \neq 0$. In this case
\begin{eqnarray}
&&z^{i} = p^{i}t + q^{i}, \qquad  i=2, \ldots ,n-1 , \\
&&z^{+} = z^{0} + z^{1} = p^{+}t + q^{+},                    \\
&&z^{-} = z^{0} - z^{1} = p^{-}t + q^{-} + \kappa^{2}A {z}(t),
\end{eqnarray}
where for $p^{+} \neq 0$
\begin{equation}
{z}(t) = 2(p^{+})^{-2} \exp(p^{+}t + q^{+}),
\qquad p^{+} p^{-} = (\vec{p})^{2}
\end{equation}
($p^{-} = 0$ for $n=2$) and for $p^{+} = 0$
\begin{equation}
{z}(t) = t^{2} \exp q^{+} , \qquad
(\vec{p})^{2} + 2 \kappa^{2} A \exp q^{+} =0.  \end{equation}
Here $(\vec{p})^{2} = \sum_{i=2}^{n-1} (p^{i})^{2}$.

For $u=0$ we have
\begin{eqnarray}
&z^{a} = p^{a}t + q^{a}, \qquad a=0, \ldots ,n-1 , \\
&\frac{1}{2}\eta_{ab}p^{a}p^{b} + \kappa^{2}A = 0.
\end{eqnarray}

{\bf Kasner-like parametrization.} Here we consider the case $u^{2} <0$,
$A \neq 0$. For $C =  -u^{2}(\vec{p})^{2} >0$  we
reparametrize the time variable
\begin{equation}
\tau = \frac{T}{\sqrt{\varepsilon}} \ln \frac
{\exp(\sqrt{C}(t-t_{0})) + \sqrt{\varepsilon}}
{\exp(\sqrt{C}(t-t_{0})) - \sqrt{\varepsilon}},
\end{equation}
where
\begin{equation}
\varepsilon \equiv  A/|A| = \pm 1, \qquad T \equiv
(2/ \kappa^{2}|A||u^{2}|)^{1/2}.
\end{equation}
We introduce new (Kasner-like) parameters
\begin{equation}
\alpha^{i} \equiv -2V^{i}_{s}p^{s}/ \sqrt{ -u^{2}(\vec{p})^{2}},
\end{equation}
where $(V^{i}_{a})  = (V^{a}_{i})^{-1}$ and the summation
parameter $s$ runs: $s = 1, \ldots , n-1$.
Then, due to relations (3.2), (3.5), (3.14)-(3.16), (3.18) and
Proposition 1 we get the following expression for the metric (1.1) [40]
\begin{equation}
g=-(\prod_{i=1}^{n} ({a_{i}}(\tau))^{2N_{i}-u_{i}}) d\tau \otimes d\tau +
\sum_{i=1}^{n} {a_{i}^{2}}(\tau)g^{(i)},
\end{equation}
where \begin{equation}
{a_{i}}(\tau) = A_{i}[\frac{\sinh(\tau \sqrt{\varepsilon} /T)}
{\sqrt{\varepsilon}}]^{2u^{i}/u^{2}}
[\frac{\tanh (\tau \sqrt{\varepsilon}/2T)}{\sqrt{\varepsilon}}]^
{\alpha^{i}},  \end{equation}
$i=1, \ldots ,n$; $A_{i}>0$ are constants and the parameters $\alpha^{i}$
satisfy the relations
\begin{eqnarray}
&&u_{i} \alpha^{i}=0, \\
&&G_{ij} \alpha^{i} \alpha^{j} = -4/u^{2}
\end{eqnarray} (see Proposition 1 and (3.30)). For the density (2.15)
we have \begin{equation}
{\rho}(\tau) = A \prod_{i=1}^{n} ({a_{i}}(\tau))^{u_{i}-2N_{i}}.
\end{equation}
We note, that $(\vec{p})^{2} = 2 \kappa^{2}|A| \prod_{i=1}^{n} A^{u_{i}}_{i}$

For $A>0$ we have an exceptional solution (3.31), (3.33), (3.34) with
the scale factors \begin{equation}
{a_{i}}(\tau) = \bar{A}_{i} \exp( \pm 2u^{i} \tau /u^{2}T),
\end{equation}
$\bar{A}_{i} >0$, $i=1, \ldots ,n$. This solution correspond to $C=0$ case
(3.17).

Remark 4. In [19] the Einstein equations (2.10), (2.11) were solved
for $A^{(\alpha)} =0$, $\alpha=1, \ldots, m$,
$\lambda^{1} \neq 0$, $\lambda^{i} = 0$,
$i >1$. The solutions [19] may be also obtained from the formulas
(3.31)-(3.34). We note that the spherically-symmetric analogue of the
solution [19] was considered in [36] (the  case $d=2$ was considered
previously in [35]). There exists an interesting special case of the
solutions [35, 36]. It is the $n$-time generalization
of the Schwarzschild solution

$g=-[(1- \frac{L}{R})^{A}]_{ab} dt^{a} \otimes dt^{b} +
(1- \frac{L}{R})^{- sp A} dR \otimes dR +
(1- \frac{L}{R})^{1 - sp A} R^{2} d \Omega^{2},$
where $L \neq 0$ and $A = (A_{ab})$ is symmetric $n \times n$ matrix,
satisfying the relation
$sp (A^{2}) + (sp A)^{2} = 2$.

We consider this solution in a separate publication.

{\bf 3.2 Two spaces with $m$-component matter}

Here we consider the following case: $n= 2$, $m \geq 2$, $A^{(\alpha)}
\neq 0$,
\begin{equation}
u^{(\alpha)} - u^{(1)} = b^{(\alpha)} u
\end{equation}
$\alpha=1, \ldots, m$,
where $u^{2} =0$, $u \neq 0$ and $b^{(\alpha)}$ are constants.

In $z$-coordinates (3.2), where the matrix $(V^{a}_{i})$ satisfies
the Proposition 1 (see the case  c) $u^{2} =0$) we have
\begin{eqnarray}
&&z^{+} = z^{0} + z^{1} = (V^{0}_{i} + V^{1}_{i})x^{i} = u_{i} x^{i}, \\
&&\Phi^{(1)} = u_{i}^{(1)} x^{i} =  \alpha_{+} z^{+} +  \alpha_{-} z^{-},
\end{eqnarray}
where $2\alpha_{+} = -< u^{1} ,u^{*}>$, $2\alpha_{-} = -< u^{1} ,u>$,
and $u^{*} = (u^{*}_{i})$ is defined by the relation :
$u_{i}^{*} x^{i} = z^{-}$
(or equivalently $< u^{*} ,u^{*}> =0$, $< u^{*} , u> = -2$).

Due to (3.37)-(3.39) the potential in (1.10) is factorized
\begin{equation}
V = {V_{+}}(z^{+}) {V_{-}}(z^{-}) , \end{equation}
where
\begin{eqnarray}
&&{V_{+}}(z^{+}) = \exp(\alpha_{+}z^{+})(\kappa^{2}A^{(1)} +
    \sum_{k=2}^{m} \kappa^{2} A^{(\alpha)}  \exp(b^{(\alpha)} z^{+}), \\
&&{V_{-}}(z^{-})  = \exp(\alpha_{-}z^{-}).
\end{eqnarray}

Let $A^{(\alpha)} > 0$,  $\alpha=1, \ldots , m$, . We consider
the $f$-gauge (2.19) with
\begin{equation}
F = e^{2f} = V.
\end{equation}
In this gauge the Lagrangian (2.20) reads
\begin{equation}
 L_{f} = -\frac{1}{2} {V_{+}}(z^{+})  \dot{z}^{+}
{V_{-}}(z^{-})\dot{z}^{-} -1. \end{equation}
In the variables
\begin{equation}
w^{\pm} = {w^{\pm}}(z^{\pm}) = \int_{z_{0}}^{z^{\pm}} dx {V_{\pm}}(x)
\end{equation}
the Lagrangian (3.44) has rather simple form
\begin{equation}
 L_{f} = -\frac{1}{2}  \dot{w}^{+} \dot{w}^{-} - 1 .
\end{equation}
The equations of motion for (3.46) give
\begin{equation}
{w^{\pm}}(t) = p^{\pm}t +  q^{\pm}.
\end{equation}
The parameters $p^{\pm}$ satisfy the energy constraint
\begin{equation}
2 E_{f} = -p^{+}p^{-} + 2 = 0 .
\end{equation}

Remark 5. It is interesting to note that the so-called $D$-dimensional
Schwarzschild-deSitter solution [44,45] may be obtained from
the considered here cosmological solution with $n = m = 2$ and
$N_{1}= 1$, $N_{2} = D -2$.

{\bf 3.3. $n$ spaces with $m$ component matter}

Now we consider the simplest case of the multicomponent matter.
We put in (1.10) $n \geq 2$, $A^{(\alpha)} > 0$,
$u^{(\alpha)} = b^{(\alpha)} u$,  $u^{2} < 0$,
where $b^{(\alpha)}$ are
constants, $\alpha=1, \ldots, m$.

In $z$-coordinates (3.2), corresponding to the case a) from the
Proposition 1, the Lagrangian (1.10) has the form (3.7) with the
potential
\begin{equation}
V_{A}= {V_{A}}(z^{0}) = \sum_{i=1}^{m} \kappa^{2}A^{(\alpha)}
\exp(2qb^{(\alpha)} z^{0}),
\end{equation}
where $q$ is defined in (3.11) ($A= (A^{(\alpha)})$).
The solutions of the equations (3.12) and (3.13) are expressed by
the formula (3.14) and the following relation
\begin{equation}
\int_{c_{0}}^{z^{0}} \, dx [2 {\cal E} + 2 {V_{A}}(x)]^{-1/2} = \pm (t-t_{0}),
\end{equation}

where $2{\cal E} = \sum_{i=1}^{n-1}(p^{i})^{2}$, and $c_{0}, t_{0}$
are constants.

\section{Quantum solutions}

The WDW equation for the model in harmonic time gauge (2.3) reads
as follows: \begin{equation}
(- \frac{1}{2 \mu}G^{ij}\partial_{i} \partial_{j} +\mu V)\Psi=0,
\end{equation}
where $\Psi={\Psi}(x)$ is "the wave function of the Universe", $V$ is the
potential (2.15),  $\partial_{i} = \partial / \partial x^{i}$ and
$G^{ij}$ are defined in (2.13). The relation (4.1) is a result of a
trivial quantization of the zero energy constraint (2.16),
written in the form $\mu E =0$. Here $\mu$ is a fundamental quantum
parameter of the theory.

 In $f$-gauge (2.19) the WDW equation should be written in the
conformally covariant form [15, 37] (such form of the WDW equation
was discussed earlier by Misner [38])
\begin{equation}
(- \frac{1}{2 \mu} {\Delta}[e^{2f} G]  + \frac{a_{n}}{\mu}
{R}[e^{2f} G]  + e^{-2f} \mu V)\Psi^{f} =0,
\end{equation}
where ${\Delta}[\hat{G}]$ and ${R}[\hat{G}]$ are the Laplace-Beltrami
operator and the scalar curvature of $\hat{G}$ respectively,
$a_{n} = (n-2)/8(n-1)$ and
\begin{equation}
\Psi^{f} = \exp[(2-n)f/2] \Psi.
\end{equation}
Without loss of generality we put $\mu =1$ below.

{\bf 4.1 One-component matter}

Here we find the quantum analogues of the classical solutions
from 3.1, i.e. we integrate the WDW equation
\begin{equation}
(- \frac{1}{2}\eta^{ab}\frac{\partial}{\partial z^{a}}
\frac{\partial}{\partial z^{b}}
+ V_{A}) \Psi =0.
\end{equation}
with the potential (3.8)-(3.10). We note, that the WDW equation for
1-component model with $n$ Ricci-flat spaces was considered previously
in [17].

a) $u^{2} <0$. In this case the WDW equation (4.4) reads
\begin{equation}
[ \frac{\partial}{\partial z^{0}} \frac{\partial}{\partial z^{0}}
- \sum_{i=1}^{n-1} \frac{\partial}{\partial z^{i}}
\frac{\partial}{\partial z^{i}} + 2 \kappa^{2}A  \exp(2qz^{0})]\Psi=0.
\end{equation}

We are seeking  solutions  of (4.5) in the following form
\begin{equation}
{\Psi}(z)=\exp(i \vec{p} \vec{z}){\Phi}(z^{0}),
\end{equation}
where $\vec{p}=(p_{1}, \ldots ,p_{n-1})$ is a constant vector
(generally from $C^{n-1}$), $\vec{z}=(z^{1}, \ldots ,z^{n-1}),
\vec{p}\vec{z} \equiv \sum_{i=1}^{n-1}p_{i}z^{i}$.
The substitution of (4.6) into
(4.5) gives
\begin{equation}
[-(\frac{\partial}{\partial z^{0}})^{2} -2\kappa^{2}A \exp(2qz^{0})]
\Phi = 2 {\cal E} \Phi,
\end{equation}
where $2 {\cal E}= \sum_{i=1}^{n-1}p_{i}^{2}$. Solving (4.7),
we get two linearly independent solutions
\begin{equation}
{\Phi}(z^{0})={B_{\nu}}(\sqrt{-2\kappa^{2}A }
q^{-1} e^{qz^{0}}), \end{equation}
where $\nu =i \sqrt{2{\cal E}}/q= i|\vec{p}|/q,$ and $B_{\nu}=I_{\nu},
K_{\nu}$ is modified Bessel function. We note, that
\begin{equation}
v=\exp{qz^{0}}= \exp(\frac{1}{2} u_{i} x^{i}) =
\prod_{i=1}^{n}a_{i}^{u_{i}/2} \end{equation}
is a  natural scale factor for the model ($a_{i}=e^{x^{i}}$).

The general solution of eq. (4.5) has the following form
\begin{equation} {\Psi}(z) =\sum_{B=I,K} \int d^{n-1} \vec{p} \,
{C_{B}}(\vec{p}){\Psi_{ \vec{p}}^{B}}(z) ,\end{equation}
where
\begin{equation}  {\Psi_{ \vec{p}}^{B}}(z) =  e^{i\vec{p}\vec{z}}
{B_{i|\vec{p}|/q}}(\sqrt{-2\kappa^{2}A } q^{-1} e^{qz^{0}}),
\end{equation} and functions $C_{B}$ ($B=I,K$) belong to an
appropriate class of (generalized) functions.

b) $u^{2} > 0$. In this case the WDW equation (4.4) reads
\begin{equation}
[- \frac{\partial}{\partial z^{1}} \frac{\partial}{\partial z^{1}}
+ \frac{\partial}{\partial z^{0}} \frac{\partial}{\partial z^{0}}
- \sum_{i=2}^{n-1} \frac{\partial}{\partial z^{i}}
\frac{\partial}{\partial z^{i}} + 2 \kappa^{2}A  \exp(2qz^{1})]\Psi=0.
\end{equation}

An analogous consideration in this case gives the general solution
(4.10)
with
\begin{equation}  {\Psi_{ \vec{p}}^{B}}(z) =   e^{i\vec{p}\vec{z}}
{B_{i{\nu}(\vec{p})}}(\sqrt{2\kappa^{2}A } q^{-1} e^{qz^{0}}).
\end{equation}
Here $\vec{p}=(p_{0}, p_{2}, \ldots ,p_{n-1})$, $\vec{z}=
(z^{0},z^{2}, \ldots ,z^{n-1})$, ${\nu}(\vec{p}) =i \sqrt{2{\cal E}}/q$,
and $2{\cal E} = p_{0}^{2} - \sum_{i=2}^{n-1}p_{i}^{2}$.

c) $u^{2} =0$ for $u \neq 0$ the WDW equation reads
\begin{equation}
[- 4 \partial_{+} \partial_{-} +
+ \sum_{i=1}^{n-1}(\frac{\partial}{\partial z^{i}})^{2}
- 2 \kappa^{2}A  \exp(z^{+})]\Psi=0,
\end{equation}
where $z^{\pm} = z^{0} \pm z^{1}$, $\partial_{\pm}=
\partial/\partial z^{\pm}$. The substitution
\begin{equation}
{\Psi}(z)=\exp(i \vec{p} \vec{z}){\Phi}(z^{+},z^{-}),
\end{equation}
with $\vec{p}=( p_{2}, \ldots ,p_{n-1})$, $\vec{z}=
(z^{2}, \ldots ,z^{n-1})$ entails
\begin{equation}
[ 4 \partial_{+} \partial_{-} + 2 {\cal E}
+ 2 \kappa^{2} A \exp(z^{+})] \Phi = 0,
\end{equation}
where $2{\cal E} = \sum_{i=2}^{n-1}p_{i}^{2}$. Introducing new
variables  $u^{0},u^{1}$, where $u^{0} \pm u^{1}= u^{\pm}$ and
\begin{equation}
u^{+} = 2 {\cal E}z^{+} + 2 \kappa^{2} A  \exp(z^{+}), \qquad
u^{-} = z^{-}
\end{equation}
we get the Klein-Gordon equation for $\Phi$ with $m^{2} =1$
\begin{equation}
( (\frac{\partial}{\partial u^{0}})^{2} -
(\frac{\partial}{\partial u^{1}})^{2} + 1)\Phi=0.
\end{equation}
It is quite obvious how to write the general solution of (4.22).

{\bf Quantum wormholes}. In the case a) $u^{2} < 0$ for $A < 0$ there
exist so-called quantum wormhole solutions of the WDW equation [34].
We present here a continuous spectrum family of these solutions. The
wave functions are following
\begin{equation} {\hat{\Psi}_{\lambda,\vec{n}}}(z)
=exp[-q^{-1}
\sqrt{-2\kappa^{2}A} e^{qz^{0}} \cosh(\lambda-q\vec{z}\vec{n})].
\end{equation}
where $\lambda \in R$ and $\vec{n}$ is unit vector:
$(\vec{n})^{2}=1$ ($\vec{n} \in S^{n-1}$). These solutions are
related with the solutions (4.11) (with $B =K$) by the formula
\begin{equation}
{\hat{\Psi}_{\lambda,\vec{n}}}(z)=\frac{1}{\pi} \int_{-\infty}^
{+\infty} dk \, {\Psi_{qk \vec{n}}}(z)e^{-ik \lambda},
\end{equation}
(such trick was suggested in [39], see also [20,41]). The solutions (4.19)
satisfy the quantum wormhole boundary conditions (in terms of
parameter $v$ (4.9): i) the wave function is exponentially damped for
large space geometries ($v \rightarrow +\infty$); ii)the  wave function
is regular when the spatial geometry degenerates ($v \rightarrow 0$).

We also note that the the functions \begin{equation}
\Psi_{m,\vec{n}}={H_{m}}(x^{0}){H_{m}}(x^{1})\exp[-\frac
{(x^{0})^{2}+(x^{1})^{2}}{2}] \end{equation}
where $H_{m}$ are Hermite polynomials, $m=0, 1, \ldots, $,

$x^{i}=(2/q)^{1/2}(-2\kappa^{2}A)^{1/4} \exp(qz^{0}/2) {f^{i}}(
\frac{1}{2}q \vec{z}\vec{n}),$ \qquad $i=0,1,$

$(f^{0}, f^{1}) = (sinh, cosh)$ are also solutions of
the WDW equation with the quantum wormhole boundary conditions.
(They are called discrete spectrum quantum wormholes.)
We note that the special cases of the solutions (4.19), (4.21)
for $u_{i} = 2N_{i}$ ($\Lambda$-term case) and $u_{i} = 2N_{i}-
2\delta_{i}^{1}$ (1-curvature case, $\lambda^{1} \neq 0$) were
considered in [41] and [20] respectively.

We also note that for b) $u^{2} > 0$ and $A > 0$ there also
exist quantum wormhole solutions. (In this case $z^{0}$ should be
replaced by $z^{1}$ in (4.19), $\vec{z}$ is defined in 3.1 b) and
$\vec{n}$ belongs to hypersphere.)

{\bf 4.2. Two spaces with $m$-component matter}

For the model from the subsection 3.2 the WDW equation (4.2)
in the $f$-gauge (3.43) has the following form ($\mu=1$)
\begin{equation}
(2 \frac{\partial}{ {V_{+}}(z^{+})\partial z^{+}}
 \frac{\partial}{ {V_{-}}(z^{-})\partial z^{-}}  + 1)\Psi=0.
\end{equation}
Indeed, for $n=2$ we have ${\Delta}[e^{2f}G]= e^{-2f}{\Delta}[G]$
, $a_{2}=0$ and $\Psi^{f} = \Psi$ (see (4.3).
In $w$-variable $w = (w^{0},w^{1})$, where $w^{0} \pm w^{1}= w^{\pm}$,
where $w^{\pm}$ are defined in (3.45), we get the Klein-Gordon
equation with $m^{2} =2$
\begin{equation}
[ (\frac{\partial}{\partial w^{0}})^{2} -
(\frac{\partial}{\partial w^{1}})^{2} + 2]\Psi=0.
\end{equation}

{\bf 4.3 $n$-spaces with $m$ component matter}

Here we present the solutions of the WDW equation (4.4) with the
potential (3.49) , i.e. quantum analogues of the classical
solutions from 3.3 are considered.

Repeating all arguments from 4.1 (case a)),we get the general
solution of (4.4)
\begin{equation} {\Psi}(z) =\sum_{*=\pm} \int d^{n-1} \vec{p} \,
{C_{*}}(\vec{p}){\Psi_{\vec{p}}^{*}}(z) ,\end{equation}
where
\begin{equation}
{\Psi_{ \vec{p}}^{*}}(z) =\exp(i\vec{p}\vec{z}){\Phi_{\vec{p}}^{*}}(z^{0}),
\end{equation}
$*=\pm$, and $\Phi^{*} = {\Phi_{\vec{p}}^{*}}(z^{0})$ are two linerly
independent solutions of the equation

\begin{equation}
[-(\frac{\partial}{\partial z^{0}})^{2} -2{V_{A}}(z^{0})]
\Phi = 2 {{\cal E}}_{\vec{p}} \Phi,
\end{equation}
with the notations for ${\cal E} = {{\cal E}}_{\vec{p}}$, $\vec{p}$,
$\vec{z}$ from 4.1.a).

We note, that for special values of parameters $A^{(\alpha)}$ and
$b^{(\alpha)}$ in the potential (3.49) the equation (4.26)
describes the quantum spin systems [42].

\section{Concluding remarks}

In this paper we investigated the multidimensional cosmological
model with $n$ ($n > 1$) Ricci-flat  spaces, filled
by $m$-component perfect fluid. In some sense, this model may be
considered as "universal" cosmological model: a lot of cosmological
models may be obtained from it under a suitable choice of parameters.
This fact may be used for "Toda-like" classification of known exact
cosmological (and spherically-symmetric) solutions of the Einstein
equations. (We note, that the  Bianchi-IX cosmological model is
described by the "Toda-like" Lagrangian (1.10) with $n=3$ and $m=6$.)

Here we integrated the Einstein and Wheeler-DeWitt equations for
some sets of parameters. But an open problem is the problem of
integrability  of the considered here model  (at classical and
quantum levels) for arbitrary values of the parameters
$m$, $n$, $N_{i}$ and $u^{(\alpha)}_{i}$. We hope to continue the
investigation of this problem in forthcoming publications.

\begin{center}{\bf Acknowledgments}   \end{center}

The authors are grateful to B. Allen,  A. A. Starobinsky
and the participants of VIII Russian Gravitational Conference
(Pushchino, 1993) for useful discussions. This work was supported in
part by the Russian Ministry of Science.

\pagebreak


\begin{thebibliography}{}

\bibitem{}{}
A. Chodos and S. Detweyler, Phys. Rev. {\bf D21} (1980) 2167.
\bibitem{}{}
P. G. O. Freund, Nucl. Phys., {\bf B209}  (1982) 146.
\bibitem{}{}
D. Sahdev, Phys. Lett. {\bf B 137}, (1984) 155. \\
E. Kolb, D. Lindley and
D. Seckel, Phys. Rev. {\bf D 30}, (1984) 1205.
\bibitem{}{}
R. Bergamini
and C. A. Orzalesi, Phys. Lett., {\bf 135B} (1984) 38.
\bibitem{}{} S.
Ranjbar-Daemi, A. Salam and J. Strathdee, Phys. Lett. {\bf 135B} (1984)
388.
\bibitem{}{}
D. Lorenz-Petzold, Phys. Lett. {\bf 148B} (1984)  43.
\bibitem{}{}
M. Gleiser , S. Rajpoot and J. G. Taylor, Ann.
Phys. (NY)  {\bf 160} (1985) 299.
\bibitem{}{}
Y.-S. Wu and Z. Wang, Phys. Rev. Lett. {\bf 57} (1986) 1978.
\bibitem{}{}
U. Bleyer and D.-E. Liebscher, Annalen d. Physik (Lpz) {\bf 44} (1987) 81.
\bibitem{}{}
G. W. Gibbons and K. Maeda, Nucl. Phys. {\bf B298} (1988) 741.
\bibitem{}{}
V. D. Ivashchuk and V. N. Melnikov, Nuovo Cimento {\bf B102} (1988) 131.
\bibitem{}{}
K. A. Bronnikov, V. D. Ivashchuk and V. N. Melnikov,
Nuovo  Cimento {\bf B102} (1988) 2O9.
\bibitem{}{}
V. A. Berezin, G. Domenech, M. L. Levinas, C. O. Lousto, and
N. D. Umerez, Gen. Relativ. Gravit. {\bf 21} (1989) 1177.
\bibitem{}{}
V. D. Ivashchuk and V. N. Melnikov, Phys. Lett. {\bf A135} (1989) 465.
\bibitem{}{}
V. D. Ivashchuk, V. N. Melnikov and A. I. Zhuk,
Nuovo Cimento {\bf B104} (1989) 575.
\bibitem{}{}
V. D. Ivashchuk and V. N. Melnikov, Chin. Phys. Lett. {\bf 7} (1990) 97.
\bibitem{}{}
U. Bleyer, D.-E. Liebscher, H.-J. Schmidt and A. I.
Zhuk, Wissenschaftliche Zeitschrift, {\bf 39} (1990) 20.
\bibitem{}{}
M. Szydlowski and G. Pajdosz, Class. Quantum Grav. {\bf 6} (1989) 1391. \\
M. Demianski and A. Polnarev, Phys. Rev. {\bf D 41}, (1990) 3003.
\bibitem{}{}
V. D. Ivashchuk, Phys. Lett. {\bf A170} (1992) 16.
\bibitem{}{}
A. Zhuk, Phys. Rev. {\bf D45} 1192 (1992).
\bibitem{}{}
A.I. Zhuk, Class. Quant. Grav., {\bf 9} (1992) 2029.
\bibitem{}{}
H. C. Lee, An Introduction to Kaluza-Klein Theories (World Scientific,
Singapore, 1984).
\bibitem{}{}
M. B. Green, J. H. Schwarz and E. Witten, Superstring Theory
(Cambridge University Press, 1987).
\bibitem{}{}
M. Toda, Theory of Nonlinear Lattices (Springer, 1981).
\bibitem{}{}
O. I. Bogoyavlensky, Commun. Math. Phys. {\bf 51} (1976) 201.
\bibitem{}{}
B. Kostant, Adv. Math. {\bf 34} (1979) 195.
\bibitem{}{}
M. A. Olshanetsky and  A. M. Perelomov, Phys. Rep. {\bf 71} (1981) 313.
\bibitem{}{}
M. Adler and P. van Moerbeke, Commun. Math. Phys. {\bf 83} (1982) 83.
\bibitem{}{}
G. W. Gibbons and C. N. Pope, Commun. Math. Phys. {\bf 66} (1979) 267.
\bibitem{}{}
G.-H. Halphen, C. R. Acad. Sc. Paris, {\bf 92} (1881) 1004.
\bibitem{}{}
L. A. Takhtajan, Teor. Mat. Fiz. {\bf 93} (1992) 330 (in English).
\bibitem{}{}
S. Chakravarty, M. J. Ablowitz and P. A. Clarkson, Phys. Rev. Lett.
{\bf 65} (1990) 1085.
\bibitem{}{}
S.-C. Lee, Phys. Lett. {\bf 149B} (1984) 98.
\bibitem{}{}
S. W. Hawking and D. N. Page, Phys. Rev. {\bf D42} (1990) 2655 .
\bibitem{}{}
K. A. Bronnikov, V. D. Ivashchuk and V. N. Melnikov,
In: Problems of Gravitation (Plenary Reports of VII Soviet Grav. Conf.,
Erevan, ErGU, 1989) p. 70.
\bibitem{}{}
S. B. Fadeev, V. D. Ivashchuk and V. N. Melnikov, Phys. Lett.
{\bf A161} (1991) 98 .
\bibitem{}{}
J. J. Halliwell, Phys. Rev. {\bf D38} (1988) 2468.
\bibitem{}{}
C. W. Misner, In Magic without Magic: John Archibald Wheeler, ed.
J. R. Klauder (Freeman, San Francisko, 1972) p. 441.
\bibitem{}{}
L. Campbell and L. Garay, Instituto de Optica,
C. S. I. C. at Madrid report, 1990.
\bibitem{}{}
V. D. Ivashchuk, submitted to Phys Lett. A.
\bibitem{}{}
V. D. Ivashchuk and V. N. Melnikov, submitted to Teor. Mat. Fiz.
\bibitem{}{}
O. B. Zaslavskii, Phys. Rep. {\bf 216} (1992) 179.
\bibitem{}{}
M. Szydlowski, Phys. Lett. {\bf A176} (1993) 22.
\bibitem{}{}
V. A. Berezin and V. A. Kuzmin, Mod. Phys. Lett. {\bf A3} (1988) 1421.
\bibitem{}{}
X. Dianyan, Class. Quantum Grav. {\bf 5} (1988) 871.


\end{thebibliography}
\end{document}